\documentstyle[astrobib,referee]{mnv2}
\newcommand{\lya}{Lyman~$\alpha$}

\def\gs{\mathrel{\raise0.35ex\hbox{$\scriptstyle >$}\kern-0.6em 
les
\lower0.40ex\hbox{{$\scriptstyle \sim$}}}}
\def\ls{\mathrel{\raise0.35ex\hbox{$\scriptstyle <$}\kern-0.6em 
ggles
\lower0.40ex\hbox{{$\scriptstyle \sim$}}}}
\def\ltorder{
\mathrel{\raise.3ex\hbox{$<$}\mkern-14mu\lower0.6ex\hbox{$\sim$}}
}
\def\gtorder{
\mathrel{\raise.3ex\hbox{$>$}\mkern-14mu\lower0.6ex\hbox{$\sim$}}
}
\def\msun{\>{\rm M_{\odot}}}

\input psfig

\title[Estimating the mass density of neutral gas at $z < 1$]
{Estimating the mass density of neutral gas at $z < 1$}

\author[Priyamvada Natarajan \& Max Pettini]
{Priyamvada Natarajan$^1$ and Max Pettini$^2$\\
$^1$Institute of Astronomy, Madingley Road, Cambridge CB3 0HA, U. K.\\
$^2$Royal Greenwich Observatory, Madingley Road, Cambridge, CB3 0EZ,
U. K.\\}
 
\begin{document}
\label{firstpage}
\maketitle

\begin{abstract}
We use the relationships between galactic H~I mass and $B$-band
luminosity determined by Rao \& Briggs to recalculate the mass density
of neutral gas at the present epoch based on more recent measures of
the galaxy luminosity function than were available to those authors.
We find $\Omega_{\rm gas}(z=0) \simeq 5 \times 10^{-4}$ in good
agreement with the original Rao \& Briggs value, suggesting that this
quantity is now reasonably secure.  We then show that, if the scaling
between H~I mass and $B$-band luminosity has remained approximately
constant since $z = 1$, 
the evolution of the
luminosity function found by the Canada-France redshift survey
translates to an increase of $\Omega_{\rm gas}$ by a factor of
$\approx 3$ at $z = 0.5 - 1$\,.  A similar value is obtained quite
independently from consideration of the luminosity function of Mg~II
absorbers at $z = 0.65$.  By combining these new estimates with data
from damped \lya\ systems at higher redshift, it is possible to
assemble a rough sketch of the evolution of $\Omega_{\rm gas}$ over
the last 90\% of the age of the universe.  The consumption of H~I gas
with time is in broad agreement with models of chemical evolution
which include the effects of dust, although more extensive samples of
damped \lya\ systems at low and intermediate redshift are required for
a quantitative assessment of the dust bias.\\
\end{abstract}

\section{Introduction}

Recently there have been major strides forward in charting the
progress of galaxy evolution. By combining the results of extensive
redshift surveys at $z <1$ with the density of star-forming galaxies
at $z > 2$ identified via the Lyman break, Madau et al. (1996) have shown
that it is possible to track the global star formation history over
most of the Hubble time.  The associated production of heavy elements
at $z > 2$ appears to be in good agreement with the typical
metallicity of the universe at these early epochs, $Z \simeq 1/13
Z_{\odot}$, as deduced from studies of damped \lya\ systems (DLAs) in
QSO spectra (Pettini et al. 1997; Lu et al. 1996).  Such low
abundances are in turn consistent with the the low rate of neutral gas
consumption implied by the observation that the mass density of gas in
damped systems is approximately constant over this redshift interval
(Storrie-Lombardi, McMahon, \& Irwin 1996; hereafter SMI96 ).  Thus
all three strands appear to lead to a roughly consistent picture of
the onset of galaxy formation in the universe.

Following these leads from $z = 2$ to the present time is difficult,
however.  On the one hand, bridging the gap from $z = 2$ to 1 in our
knowledge of the star-formation rate is stymied by the uncertainties
of photometric redshifts and the lack of distinctive spectral features
in this redshift range, at least at optical wavelengths (Connolly et al. 
1997).  On the other
hand, at $z \ltorder 1.5$ QSO absorbers become progressively less
effective for tracing the abundance of neutral gas and metals.  The
reason for this is the paucity of damped \lya\ systems at low and
intermediate redshifts, due to a combination of cosmological effects
(reduced pathlength), intrinsic evolution and possible dust bias (Pei
\& Fall 1995).  The seriousness of this shortage can be fully realised
when one considers that only two new DLAs at $z < 1.5$ with neutral
hydrogen column density $N$(H~I) $ \geq 2 \times 10^{20}$~cm$^{-2}$
have been discovered with the {\it Hubble Space Telescope} after
several years of FOS and GHRS observations (Lanzetta et al. 1997). The
full sample of known DLAs at $z < 1.5$ is still largely that
identified by Lanzetta, Wolfe, \& Turnshek (1995) from a trawl of the
{\it International Ultraviolet Explorer} data archive.

In this paper we provide new estimates of the mass density of neutral
gas between $z = 1$ and 0 in a way which does not rely on damped \lya\
systems but is based instead on the luminosity function of galaxies.
Rao \& Briggs (1993) conducted an extensive analysis of the literature
on the H~I content of galaxies of different morphological types to
determine relationships between the H~I mass ${\cal M}_{\rm H~I}$ and
the $B$-band luminosity $M_B$ for galaxies at redshift $z = 0$. They
then used these relations to derive the mass density of neutral
hydrogen at the present epoch which, expressed as a fraction of the
closure density, is $\Omega_{\rm gas}(z=0) \simeq 5 \times 10^{-4}$
(throughout this paper, we assume $H_{0} = 50$~km~s$^{-1}$~Mpc$^{-1}$,
$q_0 = 0.5$ and $\Lambda = 0$).  This estimate of $\Omega_{\rm gas}$
has been used extensively in comparisons with values at high redshift
determined from the column density distribution of DLAs (e.g. SMI96)
to show a significant decrease in the neutral gas content of the
universe (by about a factor of 6) from $z = 2 - 3$ to the present
time, presumably as a consequence of star formation.

In arriving at their measure of $\Omega_{\rm gas}$, Rao \& Briggs also
established that: {\it (i)} by far the major fraction (89\%) is from
spiral galaxies, with the remainder in irregulars, S0s, and
ellipticals; and {\it (ii)} the contribution from intergalactic H~I is
negligible. The latter conclusion appears to be confirmed by recent
results from the Arecibo Strip Survey, an unbiased 21~cm survey with a
high sensitivity to H~I of low surface density (Zwaan, Briggs, \&
Sprayberry 1997).

Since the work of Rao \& Briggs, which made use of the luminosity
function of spiral galaxies by Tammann (1986), extensive new galaxy
surveys have been published.  In \S3 of this Letter we recalculate the
value of $\Omega_{\rm gas}$ at low redshifts making use of these more
recent estimates of the local galaxy luminosity function.  By assuming
that there are no major changes in the relationship between H~I mass
and $B$-band luminosity, we then extend in \S4 the calculation of
$\Omega_{\rm gas}$ to $z = 1$ using the results of the Canada-France
redshift survey (CFRS) of Lilly et al. (1995, 1996). 
Although this assumption is yet to be tested observationally
(21~cm surveys of galaxies at even modest redshifts are beyond 
present instrumental capabilities), it turns 
out not to be unreasonable between $z = 0$ and 1
in the context of generalised 
models of cosmic chemical evolution such as those proposed 
by Pei \& Fall (1995).
In \S5 we carry out an independent check on the values of $\Omega_{\rm gas}$ at
intermediate redshifts using the statistics of Mg~II absorbers.
Finally in \S6 we compare our estimates of $\Omega_{\rm gas}$ at
different redshifts with model predictions 
and comment briefly on the extent to which current samples
of DLAs are biased by the presence of intervening dust.  Values of
$\Omega_{\rm gas}$ at different redshifts are collected in Table 1 and
plotted in Figure 1.

\section{Local estimate of $\Omega_{H~I}$}

The H~I mass contributed locally by galaxies of various morphological
types has been studied in detail by several authors (e.g. Bothun et
al. 1985; Wardle \& Knapp 1986; Tully 1988).  From the analysis of
this body of data, Rao \& Briggs (1993) derived the following
relationships between the H~I mass and the $B$-band luminosity
respectively for spirals, irregulars and E-S0s:
\begin{eqnarray}
\log {\cal M}_{\rm H~I}\,=\,({3.65\,-\,0.30\,M_B})\,\msun\\
\log {\cal M}_{\rm H~I}\,=\,({2.72\,-\,0.36\,M_B})\,\msun\\
\log {\cal M}_{\rm H~I}\,=\,({0.94\,-\,0.40\,M_B})\,\msun
\end{eqnarray}
Adopting a standard Schechter fit to
the luminosity function of each morphological type,
the H~I mass contributed is obtained by integrating,
\begin{eqnarray} 
\int {{\cal M}_{\rm H~I}}(M_B)\,\Phi({M_B})\,d{M_B}
\end{eqnarray}
over the optical luminosity function ${\Phi({M_B})}$ characterized by
$\phi^{*}$, $M_B^*$, $\alpha$ - the familiar parameters of the
Schechter function. The relations between ${\cal M}_{\rm H~I}$ and
$M_B$ in equations (1)--(3) in conjunction with the luminosity
function $\Phi(M_B)$ provide the means to compute the mass of neutral
gas contributed by each morphological type.  For a given morphological
mix of galaxies one can then compute the total mass in neutral
hydrogen; after correcting for the 25\% fraction of baryons as helium
nuclei, comparison with the local closure density $\rho_{\rm crit} =
6.94 \times 10^{10}~\msun$~Mpc$^{-3}$ finally leads to the required
$\Omega_{\rm gas}$\,.  Using this approach and the best estimates
available to them for the parameters of the luminosity functions of
spirals, irregulars and E-S0, Rao \& Briggs (1993) deduced
$\Omega_{\rm gas}(z=0) = {\rm (}4.9^{+2.0}_{-1.2}{\rm
)}\times{10^{-4}}$ (this value also includes as an upper limit the
contribution from dwarf ellipticals). The error was estimated to be
approximately 25\% primarily from the uncertainties in the
normalisations of the luminosity functions used ($\phi^{*}$ in
equation (4)), but also includes an allowance for the fact that low
surface brightness galaxies may be underrepresented in local samples
(Rao, Turnshek, \& Briggs 1995).  Fall \& Pei (1993) arrived at a
similar value of $\Omega_{\rm gas}(z=0)$ by summing up the total H~I
mass and $B$-band luminosity in the local universe out to 5 Mpc.

\section{Recent determinations of the local luminosity function}

Extensive galaxy surveys completed recently have led to new
determinations of the luminosity function of galaxies in the nearby
universe.  Here we recalculate $\Omega_{\rm gas}$ using these new data
for comparison with the initial estimate by Rao \& Briggs.

\begin{table}
\caption{Variation of the neutral gas mass density with redshift.
All values are for a $H_0 = 50$~km~s$^{-1}$~Mpc$^{-1}$, 
$q_0 = 0.5$, and $\Lambda = 0$ cosmology.}
\begin{tabular}{l|l|l|l|}
\hline
${\langle z \rangle}$ & $\Delta z$ & $\Omega_{\rm gas}$ & Reference\\ \hline
0.0 & 0.0 & ${\rm (}4.9^{+2.0}_{-1.2}{\rm )}\times{10^{-4}}$ & Rao \& Briggs 1993\\ 
0.10 & 0.0-0.2 & ${\rm (}5.6 \pm 1.4{\rm )}\times{10^{-4}}$ & this work (LCRS)\\ 
0.10 & 0.0-0.2 & ${\rm (}5.5 \pm 1.4{\rm )}\times{10^{-4}}$ & this work (ESP)\\ 
0.35 & 0.2-0.5 & ${\rm (}1.05 \pm 0.5{\rm )}\times{10^{-3}}$ & this work (CFRS)\\
0.63 & 0.5-0.75 & ${\rm (}1.35 \pm 0.6{\rm )}\times{10^{-3}}$ & this work (CFRS)\\
0.64 & 0.08-1.5 & ${\rm (}7.0 \pm 4.0{\rm )}\times{10^{-4}}$ & Lanzetta et al. 1995\\
0.65 & 0.2-1.0 & $1.7 \times{10^{-3}}$ & this work (Mg~II) \\
0.88 & 0.75-1.0 & ${\rm (}1.6 \pm 0.7{\rm )}\times{10^{-3}}$ & this work (CFRS)\\ 
1.89 & 1.5-2.0 & ${\rm (}2.05 \pm 1.2{\rm )}\times{10^{-3}}$ & Lanzetta et al. 1995\\ 
2.40 & 2.0-3.0 & ${\rm (}2.8 \pm 0.9{\rm )}\times{10^{-3}}$ & Lanzetta et al. 1995\\ 
3.17 & 3.0-3.5 & ${\rm (}3.0 \pm 1.5{\rm )}\times{10^{-3}}$ & SMI96\\ 
4.01 & 3.5-4.7 & ${\rm (}1.9 \pm 0.8{\rm )}\times{10^{-3}}$ & SMI96\\ 
\hline
\end{tabular}
\end{table}

\begin{figure*}
\centerline{
\psfig{figure=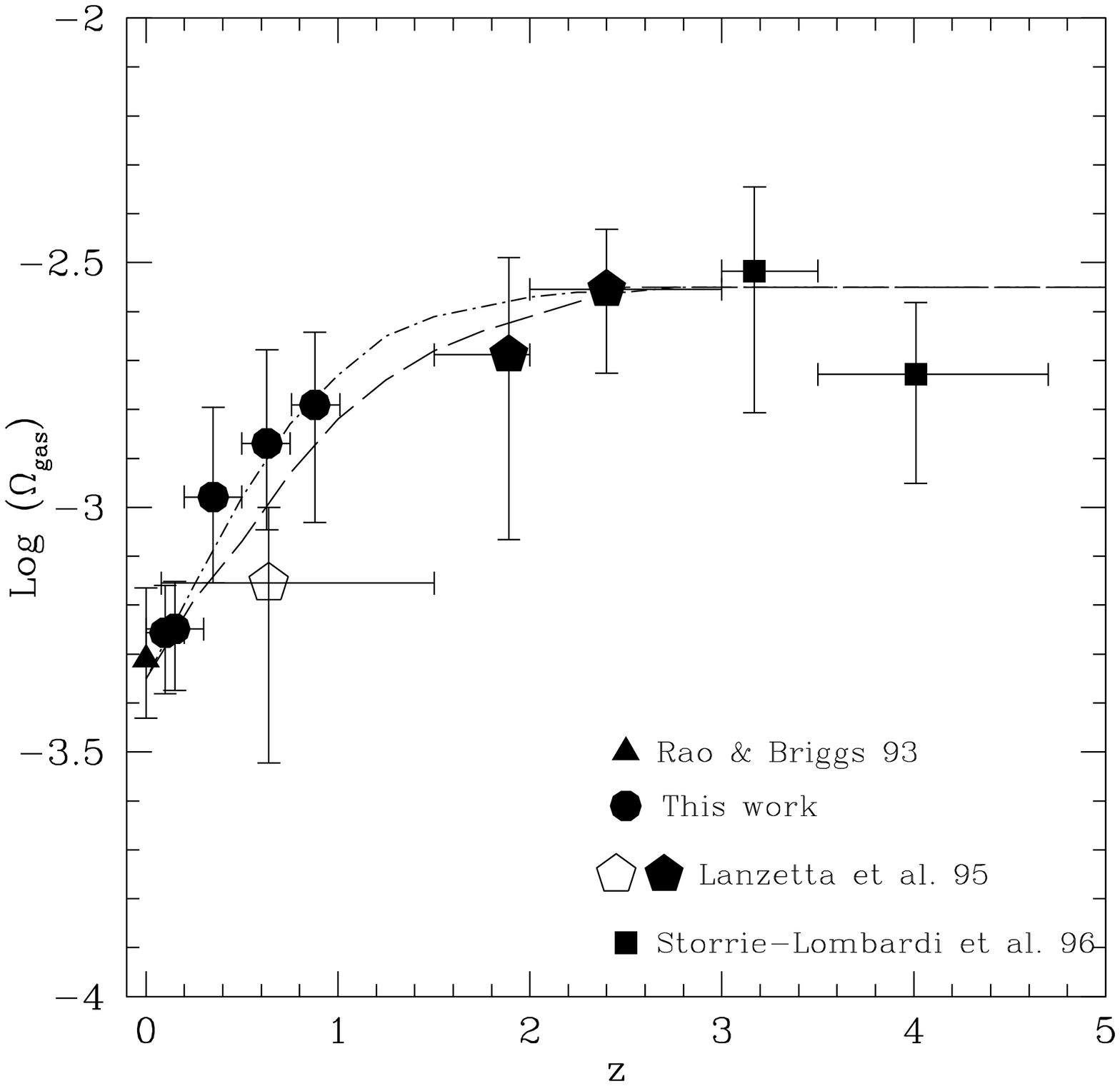,width=0.75\textwidth}}
\caption{The variation of the cosmological mass density of neutral 
gas with redshift. 
Data from the different sources indicated have all been adjusted to 
$H_0 = 50$~km~s$^{-1}$~Mpc$^{-1}$ and $q_0 = 0.5$, and include a 
correction factor of 1.3 to account for a helium fraction of 25\%\,.
Vertical error bars show estimated $1\sigma$ errors in  
$\Omega_{\rm gas}$, while the horizontal bars indicate the redshift 
interval to which each measurement applies. 
The symbols are plotted at the median redshift of each bin.
The curves show the predictions of chemical evolution models 
by Pei \& Fall (1995) with (dot-dash) and without (dash) 
inflow of metal-free gas.}
\end{figure*}

\subsection{The LCRS luminosity function}

The Las Campanas Redshift Survey (LCRS) is a magnitude limited ($-23.6
\leq M_B \leq -18.1$) $r$-band field survey of $\sim 18700$ galaxies
with average redshift $\langle z \rangle = 0.1$ (Shectman et
al. 1996).  The luminosity function of the entire sample can be fitted
with Schechter parameters: $\Phi^* = (2.38 \pm 0.13)
\times{10^{-3}}$~Mpc$^{-3}$, $M_B^* = -20.90 \pm 0.02$, and $\alpha =
-0.70 \pm 0.05$ (Lin et al. 1996).  However, these authors pointed out
that the luminosity functions of galaxies with and without emission
lines are significantly different.  Adopting an equivalent width of
[O~II]~$\lambda 3727$ $W_{\rm [O~II]} = 5$~\AA\ as the dividing line,
Lin et al.  deduced $\Phi^* = (1.63 \pm 0.13)
\times{10^{-3}}$~Mpc$^{-3}$, $M_B^* = -20.64 \pm 0.02$, and $\alpha =
-0.9 \pm 0.1$ for emission line galaxies, while non-emission line
galaxies are described by: $\Phi^* = (1.38 \pm 0.13)
\times{10^{-3}}$~Mpc$^{-3}$, $M_B^* = -20.83 \pm 0.02$, and $\alpha =
-0.3 \pm 0.1$\,.  If we assume that galaxies with $W_{\rm [O~II]} \geq
5$~\AA\ consist of spirals and irregulars in equal proportions, while
galaxies with $W_{\rm [O~II]} < 5$~\AA\ are E and S0, we find
$\Omega_{\rm gas}(z=0.1) = {\rm (}5.6 \pm 1.4{\rm
)}\times{10^{-4}}$\,.  The range quoted corresponds to changing the
spiral fraction (which dominates the contribution to $\Omega_{\rm
gas}$) from 35\% to 65\%, whereas the uncertainty arising from the
formal errors to the Schechter parameters is much smaller.

\subsection{The ESP luminosity function}

The recently completed ESO Slice Project (ESP) (Zucca et al. 1997) is
a survey of $\sim 3350$ galaxies with $b_J < 19.4$ and mean redshift
$\langle z \rangle \simeq 0.1$ distributed over $\sim 23$ square
degrees in a region near the South Galactic Pole.  As for the LCRS,
galaxies with and without emission lines yielded different fits to a
Schechter luminosity function, although Zucca et al. deduced steeper
faint-end slopes than the LCRS for both sub-samples of galaxies,
$\alpha = -1.4$ and $-1.0$ respectively.  Assuming the same
morphological mix as for the LCRS, we deduce: $\Omega_{\rm gas}(z=0.1)
= {\rm (}5.5 \pm 1.4{\rm )}\times{10^{-4}}$\,.  This is essentially
the same value as derived from the LCRS, reflecting the fact that the
major reservoirs of neutral gas are the bright spirals, rather than
the low luminosity galaxies which dominate the counts.  Similarly, the
steep rise of the luminosity function at faint magnitudes ($M_B >
-15$) recently proposed by Loveday (1997) has a negligible effect on
$\Omega_{\rm gas}$.

It is encouraging that the most recent, large scale galaxy surveys
yield values of $\Omega_{\rm gas}$ which are in good agreement with
the original estimate by Rao \& Briggs (1993).  This does suggest
that, unless a whole population of low surface brightness H~I sources
is still being missed (a possibility which seems unlikely, as
discussed by Briggs 1997), the mass density of neutral gas at the
present epoch is reasonably well established.  Note that $\Omega_{\rm
gas}(z=0) \simeq 5 \times 10^{-4}$ is approximately 13\% of
$\Omega_{\rm baryons}$ in galaxies today (e.g. Madau et al. 1996).

\section{Beyond the local luminosity function}

The CFRS has provided the first comprehensive estimate of the
luminosity function of field galaxies out to $z \sim 1$. This $I$-band
selected survey has $\sim 590$ secure redshifts with a median $\langle
z \rangle \simeq 0.56$.  The selection in the $I$-band, which in the
redshift range probed corresponds to the rest-frame $V$ and $B$ bands,
enables a comparison with local field samples to be made.  We use here
the luminosity functions which Lilly et al. (1995) derived in three
redshift intervals ($0.2 < z < 0.5$, $0.5 < z < 0.75$, and $0.75 < z<
1.0$) separately for red and blue galaxies, having divided their
sample at the spectral energy distribution of an Sbc galaxy.  The two
sub-samples such defined show markedly different evolution over the
redshift range probed.  While the luminosity function of red galaxies
remains essentially constant in both number density and luminosity,
that of blue galaxies exhibits significant evolution for $z > 0.5$.

If the relations between H~I mass and $B$-band luminosity determined by 
Rao \& Briggs (1993) also apply to galaxies up to $z = 1$ we can use the 
CFRS luminosity functions to estimate $\Omega_{\rm gas}$ in the three 
redshift bins considered by Lilly et al. (1995).
Before proceeding further, it is important to 
ask under what conditions equations (1)--(3) may have 
remained roughly constant
over the last $\approx 8$~Gyr. 
The first point to note here is that 
the $B$-band luminosity is indeed  related to the current
star formation rate, as indicated by
the correlation between H$\alpha$ luminosities and $M_B$ 
found by Tresse \& Maddox (1997) in CFRS galaxies at $z \leq 0.3$\,.
Since the time derivative of the H~I mass of a galaxy is proportional to 
the star formation rate (at least in a closed-box model), 
our assumption of no evolution in equations (1)--(3)
is satisfied only if ${\cal M}_{\rm H~I}$, and by inference $\Omega_{\rm gas}$,
decrease exponentially with time. 
It turns out that
this is approximately the behaviour of $\Omega_{\rm gas}$ from
$z \sim 1$ to the present epoch in the class of models of cosmic 
chemical evolution developed by Pei \& Fall (1995) to interpret
recent QSO absorption line measurements, as discussed in \S6 below.
Our working assumption may then be justified, 
pending an empirical determination of any redshift evolution in the 
relations between ${\cal M}_{\rm H~I}$ and $M_B$.

Brinchmann et al. (1997) have used {\it HST} WFPC2 images to study the
morphology of CFRS galaxies. Following their work, we have assumed
that the red galaxies consist of 50\% spirals and 50\% E-S0s.  For the
blue galaxies, the morphological mix changes with redshift as follows:
at $\langle z \rangle = 0.35$ 50\% of the galaxies are spirals, 40\%
irregulars, and 10\% E-S0s; at $\langle z \rangle = 0.63$ 50\% are
spirals and 50\% irregulars; and at $\langle z \rangle = 0.88$ 40\%
are spirals and 60\% irregulars.  With these weightings, equations
(1)--(3) and the CFRS luminosity functions then lead to the values of
$\Omega_{\rm gas}$ listed in Table 1.  The errors quoted, which amount
to $\approx 40-50$\%, reflect the uncertainties in the Schechter
parameters estimated by Lilly et al. (1995), added in quadrature.
Varying the spiral fractions by $\pm 15$\% increases the error on
$\Omega_{\rm gas}$ by less than 10\%,.  Despite these uncertainties,
it does appear that the evolution of the galaxy luminosity function at
$z > 0.5$ translates to higher values of the mass density of neutral
gas relative to $z = 0 - 0.2$, by factors of $\approx 2.5-3$\,.
 
\section{Comparison with Mg~II absorbers}

We can obtain an independent estimate of $\Omega_{\rm gas}$ at
intermediate redshift based on the luminosity function and
cross-section of the galaxies responsible for producing Mg~II
absorption systems in the spectra of background QSOs (Steidel,
Dickinson, \& Persson 1994).  The key point here is that selection by
Mg~II doublet with rest-frame equivalent widths $W_0 > 0.3$~\AA\ is
equivalent to selecting by neutral gas cross-section with column
density $N$(H~I)$ \gtorder 3 \times 10^{17}$~cm$^{-2}$ (Steidel 1992).
Local surveys (e.g. Zwaan et al. 1997) find that the bulk of the
contribution to the H~I mass is indeed from systems with $N$(H~I)$
\gtorder 10^{18}$~cm$^{-2}$ and with gas masses ${\cal M}_{\rm H~I}
\approx 10^{10}~\msun$.

The Mg~II absorber survey by Steidel et al. (1994) consists of 58
galaxies at $0.2 \leq z \leq 1.0$ and with median redshift $\langle z
\rangle = 0.65$\,.  The typical $B-K$ colour of the absorbing galaxies
is that of a present-day mid-type spiral, although the full sample
ranges from late-type spirals to unevolved ellipticals.  There is a
relationship, which is tighter in the $K$-band, between impact
parameter (presumably reflecting the gaseous extent of the galaxy) and
luminosity:
\begin{eqnarray}
r(L)\,=\,70\,\lgroup \frac {L}{L^*} \rgroup^{0.2}\,{\rm kpc}
\end{eqnarray}
(Steidel 1993).
We use this scaling to obtain the H~I mass as a function of $K$-band 
luminosity by modelling the neutral gas
distribution in a galaxy as an exponential disk with a mass profile of
the form:
\begin{eqnarray}
\Sigma(r)\,=\,{\Sigma_0}\,\exp - {\lgroup \frac{r}{r_d} \rgroup}
\end{eqnarray}
with a scale-length $r_d = 3.5$~kpc typical of our Galaxy and assuming that
most of the H~I mass is enclosed within a 20~kpc radius.
Combining (5) and (6) we obtain:
\begin{eqnarray}
\log {\cal M}_{\rm H~I}\,=\,({8.31\,-\,0.08\,M_K})\,\msun
\end{eqnarray}
which, together with the $K$-band luminosity
function of the Mg~II galaxies determined by Steidel et al. (1994)
then leads to
$\Omega_{\rm gas}(z=0.65) = 1.7 \times{10^{-3}}$\,.
As can be seen from Table 1, this is in good agreement
with the value deduced in \S4 from the CFRS luminosity function
at this redshift.

\section{Discussion}

We have reached two main conclusions. First, the original estimate of
the mass density of neutral gas at the present epoch by Rao \& Briggs
(1993) stands up well to the scrutiny of new large redshift surveys.
We find $\Omega_{\rm gas}(z=0.1)\,=\,(5.5 \pm 1.4) \times 10^{-4}$
from the these new surveys for $H_0 = 50$~km~s$^{-1}$~Mpc$^{-1}$, $q_0
= 0.5$, and $\Lambda = 0$. Second, under the assumption that the
relationship between H~I mass and $B$-band luminosity has not changed
significantly from $z = 0$ to 1, we find that $\Omega_{\rm gas}$ does
increase with look-back time, as expected.  Based on the luminosity
function of CFRS galaxies, we deduce values of $\Omega_{\rm gas}$ at
$z \simeq 0.5 - 1$ (between $\sim 6$ and $\sim 8$ Gyr ago in the
cosmology adopted here) which are $\approx 3$ times higher than
today's and $\approx 2$ times lower than at $z \simeq 2 - 3$ ($\sim
10$ and $\sim 11$ Gyr ago respectively).  While we do not yet know how
robust our underlying assumption is, it is encouraging that an
independent estimate of $\Omega_{\rm gas}$ at intermediate redshift,
based on the properties of galaxies selected by absorption
cross-section, is in good agreement with the CFRS values. Our approach
has provided an independent consistency check on the stellar
production rates as evidenced by the measured quantities: galaxy
counts and optical luminosities, with the exponential decline in gas
computed by Pei \& Fall (1995).

We tentatively conclude that Figure 1 gives a reasonably accurate
picture of the evolution of the neutral content of the universe over
$\sim 90$\% of its past history. 

Pei \& Fall (1995) developed models
of cosmic chemical evolution based on the measurements of $\Omega_{\rm
gas}$ at $z > 1.5$ by Lanzetta et al. (1995) and Storrie-Lombardi et
al. (1996) reproduced in Figure 1, and on the mean metallicity at
these redshifts determined by Pettini et al. (1994).  The models
include in a self-consistent way the biasing effects of dust which
become progressively more important as the average metal content of
the universe increases with time.  We show as broken lines in Figure 1
the decrease of $\Omega_{\rm gas}$ with decreasing redshift predicted
by Pei \& Fall (their Figure 4b), for a closed-box model of chemical
evolution and a model with infall of gas 
(Pei \& Fall also considered 
outflow models, but their predictions for $\Omega_{\rm gas}$
are essentially the same as those of closed-box models).  
As can be seen
from Figure 1, the models fit the data well providing a consistency
check, and, in particular, are a
good match the new values of $\Omega_{\rm gas}$ which we have derived
in this work.  In this realisation of the models, with
relatively little consumption of H~I gas through star formation until
$z \approx 1.5$, inflow does not have a major effect and, given the
uncertainties, it is not possible to distinguish models with infall of
gas from the closed-box and outflow cases.

The estimates of $\Omega_{\rm gas}$ deduced from the galaxy luminosity
functions are not greatly affected by the presence of dust.  The
difference between these values and the (lower) values obtained from
integrating the column density distribution of damped \lya\ systems
should then be a measure of the degree to which intervening dust
biases current DLA samples.  $\Omega_{\rm DLA}$ is expected to be
systematically lower than $\Omega_{\rm gas}$ because QSOs which happen
to lie behind metal enriched, and therefore dusty, galaxies (as viewed
from Earth) are preferentially missed in magnitude limited QSO
samples.  For the models reproduced in Figure 1 this effect is most
pronounced at $z < 1.5$ and indeed there are tentative indications
that $\Omega_{\rm DLA}(z=0.64)$ is only about 1/2 of $\Omega_{\rm
gas}$ at this redshift.  
However, it must be remembered that the
estimate of $\Omega_{\rm DLA}(z=0.64)$ by Lanzetta et al. (1995) is
based on only about a dozen DLAs spanning a redshift range which
corresponds to three quarters of the Hubble time, and may thus be subject 
to substantial revision. 
Indeed, very recent reports that some Mg~II absorbers
are associated with large H~I column densities suggest that the 
Lanzetta et al. value of $\Omega_{\rm DLA}$ may be an underestimate 
(Turnshek et al. 1997).
A quantitative assessment of the dust bias 
still awaits the identification
of a statistically viable
sample of intermediate redshift DLAs.

\section*{Acknowledgments}

We are grateful to Jarle Brinchmann and Ken Lanzetta for communicating 
their results in advance of publication,  
and to Simon Lilly and Mike Fall for useful discussions and comments on 
an earlier version of this work.

\end{document}